\documentclass[12pt]{article}

\usepackage{paper2e}
\usepackage{mydefs2e}
\makeatletter%  allow access to internal LaTeX commands
	\@addtoreset{equation}{section}%
\makeatother%  turn off access to internal LaTeX commands

\newcommand{\MP}{M_{\rm P}}
\newcommand{\MA}{M_{\rm AMSB}}
\newcommand{\ths}{\vartheta}

\newcommand{\lr}{\leftrightarrow}

\begin{document}

% ---------------------------------------------------------------------
% Title page
% ---------------------------------------------------------------------
\begin{titlepage}
\preprint{UMD-PP-02-046} 

\title{Weak Scale Supersymmetry\\\medskip
Without Weak Scale Supergravity}
%
% \title{Gravitino and Moduli Decoupling\\\medskip
% in Gravity-Mediated Supersymmetry Breaking}
%
% \title{Supersymmetry without Supergravity}
%
% \title{Weak Scale Supersymmetry\\\medskip
% without Weak Scale Supergravity}

\author{Markus A. Luty}

\address{Department of Physics, University of Maryland\\
College Park, Maryland 20742, USA\\
{\tt mluty@physics.umd.edu}}

\begin{abstract}
% The existence of gravity and unbroken supersymmetry implies the
% existence of the gravitino, a massless spin $\frac 32$ particle
% with couplings to matter suppressed by the Planck scale.
% Additionally, string theory and higher-dimensional supergravity also
% generically predicts moduli fields with couplings to matter
% suppressed by the Planck scale.
% It is generally believed that if supersymmetry solves the hierarchy
% problem, the gravitino mass can be no larger than of order $100$~TeV,
% with the bound saturated by anomaly-mediated supersymmetry breaking.
% % 
It is generally believed that weak scale supersymmetry implies weak scale
supergravity, in the sense that the masses of the gravitino and
gravitationally coupled moduli have masses below 100~TeV.
This paper presents a realistic framework for supersymmetry breaking in
the hidden sector in which the masses of the gravitino and gravitational
moduli can be much larger. % than $100\TeV$.
This cleanly
eliminates the cosmological problems of hidden sector models.
Supersymmetry breaking is communicated to the visible sector by
anomaly-mediated supersymmetry breaking. % or by gravity loops.
The framework is compatible with perturbative gauge coupling unification,
and 
% This framework
can be realized either in models of `warped' extra
dimensions, or in strongly-coupled four-dimensional conformal field
theories.
\end{abstract}

\end{titlepage}

% ------------------------------------------------------------------
%\section{Introduction}
% ------------------------------------------------------------------
\noindent
Supersymmetry (SUSY) is arguably the most attractive framework for 
% solving the hierarchy problem
% of the standard model of particle physics
% and
explaining the origin of electroweak symmetry breaking.
% at energies of order $100\GeV$.
SUSY automatically stabilizes the weak scale against quantum corrections
(the `hierarchy problem') and is naturally compatible with the absence
of large corrections in precision electroweak data.
If SUSY exists in nature it must be broken,
% If superpartners are discovered in upcoming experiments, and understanding
% how SUSY is broken will become the central problem in both theoretical and
% experimental particle physics.
and understanding the possible origin of SUSY breaking and its implications
for future experiments is one of the central problems in particle physics.
% It is important to remember that
% SUSY is a framework rather than a specific model:

The only truly model-independent prediction of SUSY is the existence of
superpartners of the observed particles, plus spin 0 Higgs particles
and their superpartners.
If SUSY solves the hierarchy problem, 
% solve the hierarchy problem,
% explains the origin the weak scale,
then the masses of squarks, sleptons, gauginos, Higgs fields, and Higgsinos must
be below of order $1\TeV$.

In this paper, we will focus on the gravitino, the spin $\frac 32$
superpartner of the graviton.
It is commonly thought that the gravitino mass cannot be more than of
order $100\TeV$ in models where SUSY solves the hierarchy problem,
as we now explain.
SUSY breaking gives rise to a gravitino mass of order
\beq[gravitinomass]
m_{3/2} \sim \frac{F}{\MP},
\eeq
where $F$ is the SUSY breaking order parameter
and $\MP \sim 10^{18}\GeV$ is the Planck scale.
The size of $F$ is model dependent, and depends on the strength of the
`messenger' interactions that communicate SUSY breaking to the visible
sector.
Given the fact that SUSY breaking masses in the visible sector are
between $100\GeV$ and $1\TeV$, the gravitino mass can be large if the
messenger interactions are weak.
Gravity necessarily couples the visible and hidden sectors with universal
strength, and is therefore the weakest possible messenger of SUSY breaking.
In general models of gravity mediated SUSY breaking, gravitational
contact terms give rise to superpartner masses of order $m_{3/2}$,
so in these theories $m_{3/2} \lsim 1\TeV$.

It is also possible to suppress the contact interactions between the
visible and hidden sectors \cite{RS0,compdim}.
In such models, the
supergravity contribution to supersymmetry breaking is related to the
conformal anomaly \cite{RS0,GLMR}.
The `anomaly mediated' superpartner masses are of order
$(g_{\rm SM}^2 / 16\pi^2) m_{3/2}$, which implies $m_{3/2} \lsim 100\TeV$.
% Anomaly mediation explains why the squark and slepton masses do not
% violate flavor, since gravitational interactions do not distinguish between
% different generations.
% If the visible sector is the minimal supersymmetric standard model and
% there are no additional contributions to SUSY breaking, anomaly mediation
% predicts that the slepton potential is unstable at the origin.
% There are several extensions of the minimal model that are realistic
% while preserving the attractive features of anomaly mediation \cite{amsbfix}.

String theory and higher-dimensional supergravity also
generically predict the existence of numerous moduli fields
with Planck suppressed couplings to visible matter.
In the presence of SUSY breaking, these generally get masses
of order $F / \MP \sim m_{3/2}$,
so we expect $m_{\rm moduli} \lsim 100\TeV$.

In this paper, we show that the gravitino and moduli masses can naturally
be much larger.
% 
% expectations for the gravitino and
% moduli discussed above are incorrect:
% in fact, the masses of these particles can be near the Planck scale
% in models where SUSY solves the hierarchy problem.
% % In these models, \Eqs{gravitinomass} and \eq{modulimass} are still
% % true, but the superpartner masses are suppressed compared to
% % \Eq{superpartnermass}.
% 
% This is interesting for several reasons.
This is interesting because it gives a general solution to the cosmological
problems associated with the gravitino and moduli.
Gravitinos and moduli are readily produced in inflationary reheating,
and live long because of their Planck-suppressed couplings to matter.
If the gravitino decays during or after nucleosynthesis, it can upset
the predictions of nucleosynthesis.
For $m_{3/2} \gsim 60\TeV$ the gravitino decays sufficiently rapidly to
avoid this problem \cite{GG}.
For smaller gravitino mass, we can require the reheat temperature after
inflation to be low enough to suppress the gravitino abundance.
\Refs{gravreheat} obtain $T_{\rm reheat} \lsim 10^8\GeV$ for
$m_{3/2} \sim 1\TeV$, but recent work suggests that
the gravitino production is more efficient,
requiring much lower reheat temperatures \cite{reheatnew}.
% 
% $T_{\rm reheat} \lsim 10^8\GeV$ for $m_{3/2} \sim 1\TeV$.
% To avoid problems with nucleosynthesis one requires either
% $m_{3/2} \gsim 60\TeV$ \cite{GG}, or a sufficiently low reheating
% scale so that the gravitinos are not produced \cite{gravreheat}.
Similar bounds apply to moduli, but the bounds are more
model-dependent.
% Bounds for moduli are more model-dependent, but are expected to be
% similar.

For spin 0 moduli fields, there is also the `Polonyi problem' \cite{Polonyi}.
Briefly stated, the values of the moduli fields in
the early universe differ from their present vacuum values.
This stores energy, and when this energy is released it generally reheats
the universe to a temperature too low for successful nucleosynthesis.
For $m_{\rm modulus} \gsim 100\TeV$, the reheat temperature is
$\gsim 1\MeV$, which is just enough for nucleosynthesis.

To summarize the cosmological bounds, it is fair to
say that anomaly mediation narrowly satisfies the bounds.
However, it is important that there is a class of models in which the
gravitino and moduli masses are larger and the
bounds are satisfied by a wide margin.

We now show how this can be realized in
%  in the framework of
the supersymmetric Randall--Sundrum model \cite{RS}.
This is a 5D effective field theory where the 5$^{\rm th}$
dimension is compactified on an interval of length  $\pi r$, 
realized as a $S^1 / Z_2$ orbifold.
The metric can be written
\beq
ds^2 = e^{-2 k r |\ths|} \eta_{\mu\nu} dx^\mu dx^\nu + r^2 d\ths^2,
\qquad
-\pi < \ths \le +\pi.
\eeq
The slope discontinuities in the metric at $\ths = 0$,\ $\pi$ are
due to the presence of $4D$ branes that are fixed
at the boundary by orbifold boundary conditions.
The presence of the `warp factor' $e^{-2 k r |\ths|}$
in the metric means that all physical
scales on the `IR brane' at $\ths = \pi$ are redshifted compared to the
scales on the `UV brane' at $\ths = 0$.
This model is supersymmetric with the addition of appropriate additional
fields and interactions \cite{SUSYRS}.

At energies below
the mass of the lightest gravitational Kaluza--Klein (KK) mode
% $m_{\rm KK} \sim k \avg{\om}$
the theory can be described by a 4D effective
lagrangian consisting of 4D SUGRA coupled to the radion
\cite{LS2,Bagger}:
\beq\bal
\scr{L}_{\rm 4,eff} &= -\frac{3 M_5^3}{k} \myint d^4\th
\left( \phi^\dagger \phi - \om^\dagger \om \right)
+ \myint d^4\th \left( \phi^\dagger \phi \, K_{\rm UV}
+ \om^\dagger \om \, K_{\rm IR} \right)
\\
& \quad 
+ \left[ \myint d^2 \th \left( \phi^3 \, W_{\rm UV} + \om^3 \, W_{\rm IR} 
\right) + \hc \right].
% & \quad + \myint d^4\th\, \phi^\dagger \phi \, K_{\rm UV}
% + \left( \myint d^2 \th\, \phi^3 \, W_{\rm UV} + \hc \right)
% \\
% & \quad + \myint d^4\th\, \om^\dagger \om \, K_{\rm IR}
% + \left( \myint d^2 \th\, \om^3 \, W_{\rm IR} + \hc \right).
\eal\eeq
Here $\phi$ is the conformal compensator, and the `warp factor' superfield
\beq
\om = \phi e^{-k T},
\qquad
T = \pi r + \cdots
\eeq
parameterizes the radion.
The first term contains the SUGRA and radion kinetic terms,
and the remaining terms come from \Kahler potentials
and superpotentials localized on the UV and IR branes.
% where $K_{\rm UV}$ ($K_{\rm IR}$) and $W_{\rm UV}$ ($W_{\rm IR}$)
% are \Kahler potential and superpotential
% localized on the UV (IR) brane, respectively.

We will be interested in the scenario where the visible sector fields
are localized on the IR brane and SUSY is broken on the UV brane.
Provided that there are no additional light fields
in the bulk, this automatically suppresses flavor-violating contact terms
between the visible and hidden sectors, `sequestering' the hidden sector
\cite{RS0}.
This naturally explains the absence
of flavor-changing neutral currents from superpartners.
%
% \footnote{\Ref{DGT} have shown that this does not happen in simple string
% compactifications.}
%
% In this case, we say that the hidden sector is `sequestered.'
We will be interested in the anomaly-mediated SUSY breaking (AMSB) contribution
in the visible sector.
The regulator for loops of visible sector fields must be localized
on the IR brane.
The conformal compensator for these loops is therefore $\om$, and
anomaly-mediated masses are of order $(g_{\rm SM}^2 / 16\pi^2) \MA$,
with 
$\MA = \avg{F_\om / \om}$.
The size of $\MA$ depends on the mechanism for SUSY breaking and
radius stabilization.

Radius stabilization for 5D supergravity theories
in a consistent effective field theory framework
was first achieved in \Refs{LS1,LS2}.
% In \Ref{LS2} it was shown that the radius modulus can be stabilized in
% the warped case.
However, the stabilization mechanism of \Ref{LS2} gives
$\avg{F_\om / \om} \sim \avg{F_\phi}$.
We consider a stabilization sector that gives rise to the following
additional terms in the 4D effective lagrangian:
\beq[effL54d]
\bal
\De\scr{L}_{\rm 4,eff} = & \myint d^2\th
\left( c_{\rm UV} \phi^3 + c_{\rm IR} \om^3 + \ep \phi^{3 - n} \om^n \right)
+ \hc
\\ % \nonumber\\
& \qquad -\, F_{\rm UV}^2 \left[ 1 + \hbox{\rm Goldstino\ terms} \right].
\eal\eeq
The first two terms can arise from constant superpotentials localized on
the UV and IR branes, respectively;
% the third term can arise from a massless 5D hypermultiplet with superpotentials
% localized on the UV and IR branes;
%
the 5D origin of the third term will be described below;
the last term represents the effect of SUSY breaking
on the UV brane.

In the limit $\ep \to 0$, the vacuum is at $\avg{\om} = 0$,
corresponding to infinite separation between the UV and the IR branes.
% (see \Ref{RS2}).
In this limit the fundamental scale on the IR brane is $M_{\rm IR} = M_5 \om$,
and the observations of an IR brane observer are defined relative to this
scale.
For example, the mass of the lowest KK mode is $m_{\rm KK} \sim k\om$,
which is proportional to $M_{\rm IR}$, so an IR observer sees a finite
mass gap in the KK spectrum.
% (with finite constant of proportionality).
However, the 4D Planck scale is $\MP^2 \sim M_{\rm IR}^2 / \om^2 \to \infty$,
so 4D gravity is decoupled from physics on the IR brane.
In this limit it is easy to read off the SUSY breaking from \Eq{effL54d}
because there is no mixing between the $\om$ and $\phi$
fields, and hence no supergravity corrections to the $\om$ potential.
The scale of AMSB is given by
$F_\om / \om \sim \om$, which is also proportional to $M_{\rm IR}$.
In this limit an observer on the IR brane sees SUSY broken by anomaly
mediation, even though gravity has completely decoupled!%
\footnote{I thank R. Sundrum for a discussion of this point.}
This magic is due to conformal invariance.
In the limit $\ep \to 0$ the terms that depend on $\om$ have
an exact (nonlinearly realized) conformal symmetry.
This ensures that they are independent of the conformal
compensator $\phi$.
% This can be seen from the fact that they are independent of the
% conformal compensator $\phi$.%
% %
% \footnote{Radion loops involving the $\om^3$ interaction do not
% violate conformal symmetry because it is nonlinearly realized in
% the fundamental theory \cite{Ratt}.}

To get a model with 4D gravity, we want $\avg{\om} \ne 0$.
For $\ep \ll c_{\rm UV,IR}$
and $n < 3$, the $\ep$ term gives a small shift to the vacuum:
\beq
|\avg\om|^{4-n} = \frac{n(3 - n)}{6}
\left| \frac{ \ep c_{\rm UV}}{c_{\rm IR}^2} \right| \ll 1
\eeq
with
\beq
\left| \frac{\avg{F_\om}}{\avg\om} \right| = \frac{|c_{\rm IR}|}{\MP^2}\, |\avg\om|,
\qquad
|\avg{F_\phi}| = \frac{|c_{\rm UV}|}{\MP^2} = \frac{F_{\rm UV}}{\sqrt{3} \MP},
\eeq
where $\MP^2 = M_5^3 / k$.
The radion mass is of order $\avg{F_\om / \om}$, while the mass of the
gravitino is of order $\avg{F_\phi}$.
Bulk moduli fields will also have a mass of order $\avg{F_\phi}$ provided
that the wavefunction of the lightest KK mode has sizable overlap with the
UV brane.
% \beq
% m^2_{\rm scalar} =  \frac{4 (4 - n) |c_{\rm IR}|^2}{\MP^2}
% | \avg{\om} |^2,
% \qquad
% m^2_{\rm pseudoscalar} = \frac{4 n |c_{\rm IR}|^2}{\MP^2}
% | \avg{\om} |^2.
% \eeq
Note that the order parameter for SUSY breaking on the IR brane is
parametrically suppressed (by $\avg{\om} \ll 1$) compared to $\avg{F_\phi}$.
% This can be traced to the fact that the terms in the lagrangian that
% dominate the stabilization
% % have the form
% % $\sim \myint d^4\th\, \om^\dagger \om + (\myint d^2\th \,\om^3 + \hc)$,
% % which preserves 
% preserve
% a conformal symmetry under which $\om$ has dimension 1.
%
% % We emphasize that $\avg{F_\om / \om} \ll \avg{F_\phi}$ does not
% % follow simply from $\avg{\om} \ll 1$.
% % For example, the stabilization mechanism considered in \Ref{LS2} gives
% % $\avg{\om} \ll 1$, but $\avg{F_\om / \om} \sim \avg{F_\phi}$.

We now describe the bulk interactions that give rise to the $\ep$ term in 
\Eq{effL54d}.
We add a $SU(2)$ gauge multiplet in the bulk, with 6
fundamentals with mass $m$ localized on the UV brane.
Below the scale $m_{\rm KK}$ % \sim k \avg{\om}$
this becomes a 4D $SU(2)$ gauge theory with 6 fundamentals.
If $m \lsim m_{\rm KK}$, this theory
generates a dynamical superpotential of the form \Eq{effL54d}, with
\beq
\ep \sim \left( \frac{m}{g_5^2} \right)^{3/2},
\qquad
n = \frac{4\pi^2}{k g_5^2},
\eeq
where $g_5$ is the 5D gauge coupling.
The small $m$ condition gives the constraint
\beq[smallm]
m_{3/2} \gsim \frac{\MP^2 \MA^2}{k^3 \avg{\om}^{n - 1/2}},
\eeq
where we use $g_5^2 \sim 1/k^2$ for $n \sim 1$.
% Another constraint comes from the requirement that the new gauge
% degrees of freedom are heavier than the radion mass.
% It can be checked that this is satisfied whenever $\MA$
% and $m_{3/2}$ are small compared to the compactification scale
% $m_{\rm KK}$.
There are also contact terms between the $SU(2)$ fundamentals and the SUSY
breaking sector.
These give rise to corrections to the radion potential of order
$\De V \sim m_{3/2}^2 (\ep \om^n)^{3/2}$,
which can be neglected provided
% There are also SUSY breaking contributions to the radion potential
% from integrating out the stabilization sector fields arising from contact
% terms between the $SU(2)$ fundamentals and the SUSY breaking sector.
% These are of order
% % \beq
% % \De V \sim \frac{m}{g_5^2}\, m_{3/2}^2 \om^{3n/2},
% $\De V \sim m_{3/2}^2 (\ep \om^n)^{3/2}$,
% % \eeq
% and can be neglected provided
\beq[DeltaVsmall]
m_{3/2}^2 \lsim \MA m_{\rm KK}.
\eeq
The inequalities \Eqs{smallm} and \eq{DeltaVsmall} justify the use of the
4D effective field theory above for stabilizing the radion.

We now investigate the conditions under which $\avg{F_\om / \om}$ dominates
SUSY breaking for fields on the IR brane.
First we consider SUSY breaking from SUGRA loops.
%  such as those in Fig.~1.
In a 4D effective theory, the loop is dominated by a quadratically
divergent contribution from 4D momenta above $m_{3/2}$.
(We are restricting attention to the case $m_{3/2} \lsim m_{\rm KK}$
where we can treat the gravitino in the 4D effective theory.)
This gives
\beq[radstab4d]
\De m_{\rm scalar}^2 \sim \frac{m_{3/2}^2 \La_{\rm 4D}^2}{16\pi^2 \MP^2},
\eeq
where $\La_{\rm 4D}$ is the UV cutoff in the 4D theory.
In the 5D theory the integral is cut off because in
position space the SUGRA propagators must extend from the IR brane to the
UV brane in order to communicate SUSY breaking.
Because the gravitino loop cannot shrink to zero size, so there is no UV
divergence \cite{RS0}.
% We can make this more precise as follows.
We can understand the size of this finite loop contribution as follows.
For 4D loop momenta $p_4 \lsim m_{\rm KK}$, the
graviton loop behaves as in the 4D theory.
For $p_4 \gsim m_{\rm KK}$ the brane-to-brane propagator falls off
as $e^{-|p_4|/m_{\rm KK}}$, so the result is given by \Eq{radstab4d}
with $\La_{\rm 4D} \sim m_{\rm KK}$.
The AMSB contribution is larger than the gravity loop contribution
provided that%
\footnote{The case where this inequality is saturated may be interesting.
\Ref{GR} did the calculation for a flat extra
dimension and obtained a negative mass-squared.
The calculation has not been done for the warped case.}
\beq[SUGRAsmall]
% m_{3/2} \lsim \frac{\MP \MA}{4\pi m_{\rm KK}}.
\frac{m_{3/2}}{\MA} \lsim \frac{\MP}{4\pi m_{\rm KK}}.
% k^2 \lsim \frac{c_{\rm IR}}{4\pi \MP \avg{\om}}.
\eeq
% We note that the case where this inequality is saturated is very interesting.
% First of all, it is flavor-blind because it arises entirely from gravitational
% interactions.
% This contribution can be the same size as the anomaly-mediated contribution
% for a special (but reasonable) choice of parameters \cite{RS0}.
% $\ldots$

SUSY breaking can also be communicated to the visible sector through loops
of $SU(2)$ gauge fields from the stabilization sector.
The $SU(2)$ gauge fields couple to observable fields via
flavor-violating interactions on the
IR brane of the form
\beq[IRcontact]
\De\scr{L}_{\rm IR} \sim \myint d^4\th\, \om^\dagger \om
\frac{1}{M_5^5} Q^\dagger Q W^{\al} \si^\mu_{\al\dot\al} \partial_\mu
\bar{W}^{\dot\al},
\eeq
where $Q$ is a visible sector field and
$W_\al$ is the $SU(2)$ field strength.%
\footnote{In \Eq{IRcontact} we define all fields to have vanishing conformal weight.}
The $SU(2)$ gaugino gets a mass from the coupling to the radion of order
$\avg{F_T / T} \sim \avg{F_\phi / \ln\om}$ \cite{RMSB},
which gives rise to flavor-violating visible scalar masses.
The condition that these are smaller than experimental bounds gives
\beq[SUSYstabsmall]
% \frac{m_{3/2}^2}{\MA^2}
% \lsim 10^{-5} \left( \frac{\MP}{k} \right)^{10/3}
% \frac{(\ln\om)^2}{\om^5}.
\frac{m_{3/2}}{\MA}
\lsim 10^{-2} \left( \frac{\MP}{k} \right)^{5/3}
\frac{\ln\avg{\om}}{\avg{\om}^{5/2}}.
\eeq
Provided that \Eqs{smallm}, \eq{DeltaVsmall}, \eq{SUGRAsmall},
and \eq{SUSYstabsmall} are satisfied, SUSY breaking for fields localized on
the IR brane is dominated by anomaly-mediated SUSY breaking from
$\avg{F_\om / \om}$.

We now discuss the phenomenology of this model.
SUSY breaking is communicated to the visible sector by anomaly mediation.
This naturally gives a flavor-blind squark masses and therefore explains
the absence of flavor-changing neutral currents from squark mixing.
If the visible sector is the minimal supersymmetric standard model, the
slepton mass-squared terms are negative \cite{RS0}.
There are a number of proposals in the literature for non-minimal
models that give a realistic spectrum while preserving the
attractive features of anomaly mediation \cite{AMSBfix}.
% Alternatively, we can assume that the inequality \Eq{SUGRAsmall} is saturated,
% in which case gravity loops give rise to flavor-blind scalar masses
% $\ldots$

In order to preserve the success of perturbative gauge coupling unification,
we require that $M_5 \om \gsim M_{\rm GUT} \sim 10^{16}\GeV$.
The largest allowed value for the gravitino mass in this scenario is
$m_{3/2} \sim 10^9\GeV$, with
$m_{\rm KK} \lsim 10^{13}\GeV$ and
$\avg\om \sim 10^{-2}$.
For $k \sim M_5$ the largest value is $m_{3/2} \sim 10^6\GeV$.
If one is willing to give up perturbative gauge coupling unification,
higher values of $m_{3/2}$ can be achieved.

The duality between 5D anti De Sitter space and 4D conformal field theory
(CFT) \cite{AdSCFT,AdSCFTphen} suggests that there is another realization
of this scenario where the role of the
5D warped bulk is played by a strongly coupled 4D CFT.
\Ref{compdim} showed that 4D CFT's with no known 5D `dual'
description can lead to a sequestered hidden sector and anomaly mediated
SUSY breaking.
Along these lines, we now show that the qualitative features of the 5D
models described above can be realized in a 4D theory based on a
strongly-coupled CFT.
This gives additional insight into the mechanism described above.

Consider a 4D $SU(2)$ SUSY gauge theory with 8 fundamentals $P$.
This theory is asymptotically free in the UV, and approaches a strongly
coupled conformal fixed point in the IR \cite{SeibergCFT}.
We add to this theory a superpotential
\beq
W = \la P^4 + \ka P^2.
\eeq
At the fixed point, the dimension of $P$ is $\frac 34$, so $\la$ is dimensionless
and $\ka$ has dimension $\frac 32$.
%
% \footnote{The dimensions of chiral composite operators are given by the sum
% of the dimensions of the constituent fields \cite{Mack}.}
%
We can parameterize the moduli space by
\beq
PP = \bordermatrix{& 2 & 6 \cr
2 & \pmatrix{0 & X \cr -X & 0 \cr} & Y \cr
6 & -Y^T & \scr{O}(Y^2 / X) \cr},
% \qquad
% \ep = \pmatrix{0 & 1 \cr -1 & 0 \cr},
\eeq
and expand about $X \ne 0$, $Y = 0$.
The (nonlinearly realized) conformal symmetry implies that the 
effective field theory below the scale $X$ has \Kahler terms \cite{compdim}
\beq
\De\scr{L}_{\rm eff} &= 
\myint d^4\th\, \phi^\dagger \phi
(X^\dagger X)^{2/3} \left[ 1 + \scr{O}(|Y|^2 / |X|^2)
\right]
\\
&= \myint d^4\th \hat{X}^\dagger \hat{X}
\left[ 1 + \scr{O}(|\hat{Y}|^2 / |\hat{X}|^2)
\right],
\eeq
where $\hat{X} = \phi X^{2/3}$, $\hat{Y} = \phi Y / X^{1/3}$.
The $\phi$ dependence can be completely scaled away because of
% The \Kahler terms are independent of
% the superconformal compensator $\phi$ because of
conformal invariance.
The effective superpotential terms are
\beq
\De\scr{L}_{\rm eff} = 
\myint d^2\th \left[ \la \hat{X}^3 + \ep \phi^{3 - n} \hat{X}^n 
+ (\hbox{$Y$\ dependent}) \right] + \hc,
\eeq
with $n = \frac 32$.
This is precisely the same effective lagrangian we obtained in the 5D case
with the identifications $\hat{X} \lr \MP \om$,
$\la \lr c_{\rm IR} / \MP^3$,
$\ka \lr \ep / \MP^n$.

The composite fields $\hat{Y}$ correspond to the fields localized on the
IR brane in the 5D model.
Loops of $\hat{Y}$ fields are cut off at a scale proportional to $\hat{X}$,
so the anomaly-mediated contribution to their masses is
determined by $\avg{F_{\hat{X}} / \hat{X}} \ll \avg{F_\phi}$, as in
the 5D model.
% This shows that the basic features of the model above can be simply realized
% in a 4D CFT.
The suppression of anomaly-mediated contribution proportional to $F_\phi$
in both frameworks is closely related to conformal invariance.

A complete model requires that the standard model gauge bosons
are composite.%
\footnote{Composite gauge bosons are known to emerge in simple SUSY
gauge theories \cite{SeibergCFT}.}
This is compatible with perturbative unification if the compositeness
scale is above the GUT scale.
Sequestering also requires that there are no unbroken
global (\eg flavor) symmetries in
the visible sector \cite{compdim}.
\Ref{NS} gives an interesting mechanism for generating the observed
flavor structure in this type of scenario.
We conclude that the ingredients for a fully realistic model can be
realized in 4D CFT's.
Our ability to construct explicit realistic 4D models is
limited mainly by our poor understanding of strongly coupled
superconformal field theories.

% To make a fully realistic 4D model of this type, one would have to generate the
% standard model gauge bosons as composites of a CFT.
% In such a model, the standard model gauge fields are strongly coupled at the
% compositeness scale, but this is compatible with unification if the compositeness
% scale is above the GUT scale.
% Constructing models of this type is beyond the scope of the
% present work.

We
have restricted attention to the regime $m_{3/2} \lsim m_{\rm KK}$, where
the gravitino can be treated in the 4D effective field theory.
There appears to be no fundamental reason that we cannot have
$m_{3/2} \gg m_{\rm KK}$, and this is presently under investigation.

In conclusion, we have shown that the masses of the gravitino and gravitational
moduli can be much larger than
% expectations based on gravity-mediated
% (and anomaly-mediated) supersymmetry breaking,
the weak scale,
without upsetting the major successes of supersymmetry:
a solution to the hierarchy problem and the
successful predictions of perturbative gauge coupling unification.
The only assumption about the moduli is that 
SUSY breaking in the moduli has gravitational strength.
% This mechanism is very general and applies to all bulk fields that have
% significant overlap with the UV brane.
This may be of particular interest in superstring/M theory, where
quasi-realistic compactifications have many moduli that give rise to
cosmological difficulties.

\renewcommand{\baselinestretch}{0.5}  % line spacing
\makeatletter
\setlength{\textheight}{8.6in}%8.4in
\makeatother

\newpage

\section*{Acknowledgements}
I thank N. Arkani-Hamed,
R. Sundrum,  M. Schmaltz, Y. Shirman,
and N. Weiner for constructive skepticism and useful conversations.
I also thank E. Pont\`on, R. Sundrum, and N. Weiner for critically reading
the manuscript.
This work was supported by NSF grant PHY-0099544.

\vskip -1in
% ---------------------------------------------------------------------


\begin{thebibliography}{99}
% ---------------------------------------------------------------------
\bibitem{RS0}
L. Randall, R. Sundrum, {\em Nucl.\ Phys.}\ {\bf B557} (1999) 79
[{\tt hep-th/9810155}].

\bibitem{compdim}
M.A.~Luty, R.~Sundrum,
{\it Phys.\ Rev.}\ {\bf D65}, 066004 (2002)
[{\tt hep-th/0105137}];
{\tt hep-th/0111231}.
%%CITATION = HEP-TH 0111231;%%

\bibitem{GLMR}
G.F. Giudice, M.A. Luty, H. Murayama, R. Rattazzi,
{\em JHEP} {\bf 9812} (1998) 027 [{\tt hep-ph/9810442}].

%\cite{Gherghetta:1999sw}
\bibitem{GG}
T.~Gherghetta, G.F.~Giudice, J.D.~Wells,
%``Phenomenological consequences of supersymmetry with anomaly-induced  masses,''
{\it Nucl.\ Phys.}\  {\bf B559}, 27 (1999)
[{\tt hep-ph/9904378}].
%%CITATION = HEP-PH 9904378;%%

\bibitem{gravreheat}
J. Ellis, G.B. Gelmini, J.L. Lopez, D.V. Nanopoulos, and S. Sarkar,
{\em Nucl.\ Phys.}\ {\bf B373}, 399 (1992).
For an updated analysis, see
M.~Kawasaki, K.~Kohri, T.~Moroi,
%``Radiative decay of a massive particle and the non-thermal process in  primordial nucleosynthesis,''
{\em Phys.\ Rev.}\ {\bf D63}, 103502 (2001)
[{\tt hep-ph/0012279}].

\bibitem{reheatnew}
R. Kallosh, L. Kofman, A Linde, A.V. Proeyen, {\em Phys.\ Rev.}\ {\bf D61}
103503 (2000) [{\tt hep-th/9907124}].

\bibitem{Polonyi}
J. Polonyi, Budapest preprint KFKI-93 (1977);
G.D. Coughlan, R. Holman, P. Ramond, G.G. Ross,
{\it Phys. Lett.} {\bf B140} 44 (1984);
B. de Carlos, J.A. Casas, F. Quevedo, and E. Roulet,
{\it Phys. Lett.} {\bf B318} 447 (1993) [{\tt hep-ph/9308325}];
T. Banks, D. B. Kaplan, and A. Nelson,
{\it Phys.\ Rev.} {\bf D49} 779 (1994) [{\tt hep-ph/9308292}].


\bibitem{RS}
L.~Randall, R.~Sundrum,
%``A large mass hierarchy from a small extra dimension,''
{\em Phys. Rev. Lett.} {\bf 83} (1999) 3370 [{\tt hep-ph/9905221}].

\bibitem{SUSYRS}
M. Cvetic, H. Lu, C.N. Pope, {\tt hep-th/0002054};
R. Altendorfer, J. Bagger, D. Nemeschansky, 
{\it Phys.\ Rev.}\ {\bf D63}, 125025 (2001)
[{\tt hep-th/0003117}];
T. Gherghetta, A. Pomarol,
{\it Nucl.\ Phys.}\ {\bf B586}, 141 (2000)
[{\tt hep-ph/0003129}];
N. Alonso-Alberca, P. Meessen, T. Ortin, {\em Phys. Lett.}
{\bf B482} (2000) 400, {\tt hep-th/0003248};
A. Falkowski, Z. Lalak, S. Pokorski, 
{\it Phys.\ Lett.}\ {\bf B491}, 172 (2000) [{\tt hep-th/0004093}];
E. Bergshoeff, R. Kallosh, A. Van Proeyen,
{\em JHEP} 0010 (2000) 033, [{\tt hep-th/0007044}];
M. Zucker, {\it Phys.\ Rev.}\ {\bf D64}, 024024 (2001)
[{\tt hep-th/0009083}].

\bibitem{LS2}
M.A. Luty, R. Sundrum,
{\it Phys.\ Rev.}\ {\bf D64}, 065012 (2001)
[{\tt hep-th/0012158}].

\bibitem{Bagger}
J.~Bagger, D.~Nemeschansky, R.-J.~Zhang,
{\it JHEP} {\bf 0108}, 057 (2001)
[{\tt hep-th/0012163}]. 

\bibitem{LS1}
M.A.~Luty, R.~Sundrum,
{\em Phys.\ Rev.}\ {\bf D62} 035008 (2000)
[{\tt hep-th/9910202}].


\bibitem{GR}
%\cite{Gherghetta:2001sa}
T.~Gherghetta, A.~Riotto,
%``Gravity-mediated supersymmetry breaking in the brane-world,''
{\it Nucl.\ Phys.}\ {\bf B623}, 97 (2002)
[{\tt hep-th/0110022}].
%%CITATION = HEP-TH 0110022;%%

\bibitem{RMSB}
Z.~Chacko, M.A.~Luty,
%``Radion mediated supersymmetry breaking,''
{\it JHEP} {\bf 0105}, 067 (2001)
[{\tt hep-ph/0008103}].
%%CITATION = HEP-PH 0008103;%%

\bibitem{AMSBfix}
A.~Pomarol, R.~Rattazzi,
%``Sparticle masses from the superconformal anomaly,''
{\em JHEP} {\bf 9905} 013 (1999) [{\tt hep-ph/9903448}];
Z. Chacko, M.A. Luty, I. Maksymyk, E. Pont\`on,
{\em JHEP} {\bf 0004} 001 (2000)
[{\tt hep-ph/9905390}];
E. Katz, Y. Shadmi, Y. Shirman {\em JHEP} {\bf 9908} 015 (1999)
[{\tt hep-ph/9906296}];
K.I. Izawa, Y. Nomura, T. Yanagida, {\em Prog. Theor. Phys.} {\bf 102}
1181 (1999) [{\tt hep-ph/9908240}];
M. Carena, K. Huitu, T. Kobayashi, {\em Nucl. Phys.} {\bf B592} 164 (2000)
[{\tt hep-ph/0003187}];
B.C.~Allanach, A.~Dedes,
{\em JHEP} {\bf 0006} 017 (2000) [{\tt hep-ph/0003222}];
I. Jack, D.R.T. Jones, {\em Phys. Lett.} {\bf B491} 151 (2000)
[{\tt hep-ph/0006116}];
D.E. Kaplan, G.D. Kribs, {\em JHEP} 0009 048 (2000) [{\tt hep-ph/0009195}].
N.~Arkani-Hamed, D.E.~Kaplan, H.~Murayama, Y.~Nomura,
%``Viable ultraviolet-insensitive supersymmetry breaking,''
{\em JHEP} {\bf 0102} 041 (2001)
[{\tt hep-ph/0012103}];
Z.~Chacko and M.A.~Luty,
%``Realistic anomaly mediation with bulk gauge fields,''
{\tt hep-ph/0112172};
A.E.~Nelson and N.J.~Weiner,
%``Gauge/anomaly Syzygy and generalized brane world models of  supersymmetry breaking,''
{\tt hep-ph/0112210}.
%%CITATION = HEP-PH 0112172;%%

\bibitem{AdSCFT}
J.~Maldacena,
{\it Adv.\ Theor.\ Math.\ Phys.}\ {\bf 2} 231 (1998) [{\tt hep-th/9711200}];
S.S.~Gubser, I.R.~Klebanov, A.M.~Polyakov,
{\it Phys.\ Lett.}\ {\bf B428} 105 (1998) [{\tt hep-th/9802109}];
E.~Witten, {\it Adv.\ Theor.\ Math.\ Phys.}\ {\bf 2}
253 (1998) [{\tt hep-th/9802150}].


\bibitem{AdSCFTphen}
H. Verlinde, {\em  Nucl. Phys.} {\bf  B580} 264 (2000)
[{\tt hep-th/9906182}];
J. Maldacena, unpublished remarks;
%H. Verlinde, talk at ITP Santa Barbara conference
%`New Dimensions in Field Theory and String Theory,'
%{\tt http://www.itp.ucsb.edu/online/susy c99/verlinde};
E. Witten, ITP Santa Barbara conference
`New Dimensions in Field Theory and String Theory,',
{\tt http://www.itp.ucsb.edu/online/susy
c99/discussion};
S. S. Gubser, {\em Phys.\ Rev.}\ {\bf D63} 084017 (2001)
[{\tt hep-th/9912001}];
E.~Verlinde, H.~Verlinde, {\em  JHEP} {\bf 0005} 034 (2000)
[{\tt hep-th/9912018}];
N. Arkani-Hamed, M. Porrati, L. Randall, {\it JHEP} {\bf 0108} 017 (2001)
[{\tt hep-th/0012148}];
R. Rattazzi, A. Zaffaroni,
{\em JHEP} {\bf 0104} (2001) 021
[{\tt hep-th/0012248}];
M. Perez-Victoria, {\it JHEP} {\bf 0105} 064 (2001)
[{\tt hep-th/0105048}].

\bibitem{SeibergCFT}
N.~Seiberg,
%``Electric - magnetic duality in supersymmetric nonAbelian gauge theories,''
{\it Nucl.\ Phys.}\ {\bf B435}, 129 (1995)
[{\tt hep-th/9411149}].

\bibitem{NS}
A.~E.~Nelson, M.~J.~Strassler,
%``Suppressing flavor anarchy,''
{\em JHEP} {\bf 0009}, 030 (2000)
[{\tt hep-ph/0006251}].
%%CITATION = HEP-PH 0006251;%%

\end{thebibliography}
\end{document}